\documentstyle[12pt,epsf]{article}

\makeatletter
\@addtoreset{equation}{section}

\def\ben{\begin{equation}}
\def\een{\end{equation}}
\def\bena{\begin{eqnarray}}
\def\eena{\end{eqnarray}}

\def\b1{e^0}

\input amssym.def
\input amssym.tex

\newcommand{\nn}{\nonumber}
\newcommand{\vs}[1]{\vspace*{#1}}
\newcommand{\hs}[1]{\hspace*{#1}}

\newcommand{\p}{\partial}
\newcommand{\Half}{\frac12}
\newcommand{\unit}{\hbox to 3.8pt{\hskip1.3pt \vrule height 7.4pt
    width .4pt \hskip.7pt \vrule height 7.85pt width .4pt \kern-2.4pt 
    \hrulefill \kern-3pt \raise 3.7pt\hbox{\char'40}}}

\def\href#1#2{#2}

\begin{document}

\begin{titlepage}
\title{
\vspace{-20mm}
\hfill\parbox{4cm}{
{\normalsize DAMTP-2000-65}\\[-5mm]
{\normalsize YITP-00-36}\\[-5mm]
{\normalsize\tt hep-th/0007019}
}
\\
\vspace{15mm}
Non-Linear Electrodynamics in Curved Backgrounds
}
\author{
{}
\\
Gary W.\ 
{\sc Gibbons}${}^1$\thanks{{\tt G.W.Gibbons@damtp.cam.ac.uk}}
\hs{5mm}
and 
\hs{5mm}
Koji {\sc Hashimoto}${}^2$\thanks{{\tt hasimoto@yukawa.kyoto-u.ac.jp}}
\\[7pt]
${}^1$ {\it DAMTP, CMS, University of Cambridge}, \\
{\it Wilberforce Road, Cambridge CB3 OWA, UK}\\[7pt]
${}^2$ {\it Yukawa Institute for Theoretical Physics,}\\
{\it Kyoto University, Kyoto 606-8502, Japan}\\[5pt]
}
\date{\normalsize July, 2000}
\maketitle
\thispagestyle{empty}


\begin{abstract}
\normalsize\noindent 
We study non-linear electrodynamics in curved space from the viewpoint 
of dualities. After establishing the existence of a topological bound
for self-dual configurations of Born-Infeld field in curved space, 
we check that the energy-momentum tensor vanishes. These properties
are shown to hold for general duality-invariant non-linear 
electrodynamics. We give the dimensional reduction of Born-Infeld
action to three dimensions in a general curved background admitting a
Killing vector. The $SO(2)$ duality symmetry becomes manifest but
other symmetries present in flat space are broken, as is U-duality
when one couples to gravity. We generalize our arguments on duality
to the case of  $n$ $U(1)$ gauge fields, and present a new Lagrangian
possessing $SO(n)\times SO(2)_{\rm elemag}$ duality symmetry. Other
properties of this model such as Legendre duality and  enhancement of
the symmetry by adding dilaton and axion, are studied. We extend our
arguments to include a background $b$-field in the curved space, and
give new examples including almost K\"ahler manifolds and Schwarzshild 
black holes with a $b$-field. 
\end{abstract}

\end{titlepage}

\section{Introduction}

Non-linear electrodynamics has been studied for a long time.
M.\ Born and L.\ Infeld \cite{BI} introduced a version of it 
which has received renewed attention since it 
has turned out to play an important role in the  development in
string theory. The particular non-linear electrodynamics of
the Born-Infeld Lagrangian describes low energy physics on D-branes
which are intrinsically non-perturbative solitonic objects in string
theory.  
The introduction of  D-brane has expanded 
the  possibilities for  constructing realistic 
models in string theory, and one of those possibilities is
the  so-called
``brane-world'' scenario which assumes that our spacetime is 
a worldvolume of the (D-)branes. Therefore this scenario naturally
introduces  non-linear electrodynamics into gauge theories.

Since string theory contains gravity as its fundamental excitation, 
it is plausible that  the low energy effective action on the D-brane
incorporates  gravity. However, the effective action for  D-brane was
derived initially only in a  flat
background. Therefore the precise form of  non-linear
electrodynamics on the D-brane in arbitrary gravity background,
while probably given by the covariant form of the 
Born-Infeld Lagrangian \cite{Zeid}, is still not 
yet known with certainty, especially in the case of more than one
$U(1)$ gauge field. Recently, in the supergravity context, progress on
the construction of D-brane solutions possessing non-flat worldvolumes
has been reported \cite{Brecher, Papadopoulos, Kaya}. These are called
Ricci-flat branes. If we accept  the brane-world scenario, then surely
we have curved spacetime as our universe, thus we need to understand
better the relation between non-linear electrodynamics
and gravity .  

In this paper, we shall study  classical non-linear electrodynamics in
curved background. Since string theory is believed to be controlled by
(non-)perturbative duality symmetry and in particular S-duality
may be realized as a {\it non-linear Legendre (electric-magnetic)
  duality} on the D3-brane 
worldvolume theory, we concentrate on the duality properties of the
non-linear electrodynamics in curved backgrounds. In particular,
self-dual configurations of the gauge field are closely related to
Legendre duality, therefore we study mainly the ``instanton"
configurations. In the string theory context,  self-dual (or
anti-self-dual) configurations are concerned with stable (BPS)
brane configurations such as bound states of branes.  

We see that the self-dual configurations in the curved gravity
background are involved with various physical requirements which
constrain the explicit form of the Lagrangian of the non-linear
electrodynamics. Our aim is to find Lagrangian(s) which possesses
desirable properties concerning the self-duality and self-dual
configurations. 

Another motivation of this paper  concerns  non-commutative
geometry. String theory in a  background NS-NS two-form $b$-field
gives rise to  exotic worldvolume physics on the D-brane, involving
``non-commutative" spacetimes. The non-commutativity of the
worldvolume can be introduced through a $*$-product in the
multiplication of 
functions on that space. This exotic theory is, as shown in Ref.\
\cite{SW}, equivalent with the ordinary Born-Infeld type nonlinear
electrodynamics in a  background $b$-field. The equivalence involves
 non-linear electrodynamics in an essential way, one must go beyond
Maxwell's  linear electrodynamics. 

 From this point of view, we introduce the $b$-field in  nonlinear
electrodynamics on a curved background. Consequently the $b$-field is
no longer strictly constant. It may be covariantly constant,
it may have constant magnitude or it may merely
tend to a constant field near infinity.
 We shall call such  manifolds with a  non-trivial $b$-field
``non-commutative manifolds''. The analysis of the self-dual
configurations of the gauge fields is extended to include
the case of non-commutative manifolds in this sense. 

The organization of this paper is as follows. First in the next
section we shall consider the Abelian Born-Infeld theory in the curved
background, and see that the notion of the topological bound for
instantons survives in the case of curved gravity background. In Sec.\ 
\ref{sec:abelian}, we extend the analysis to general non-linear
electrodynamics, and show that the properties concerning the
instantons can be derived from the duality invariance condition 
of the system. Sec.\ \ref{sec:higher} is devoted to the
study of the case of several gauge fields. There are many versions of 
extension of the Abelian Born-Infeld theory, however we propose a new
Lagrangian which possesses intriguing duality properties. 
In Sec.\ \ref{sec:property}, subsequently we study various properties
of non-linear electrodynamics such as Hamiltonian formalism, Legendre
duality, symmetry enhancement by adding dilaton and axion, and
dimensional reduction in curved space.
Sec.\ \ref{sec:curved} includes many examples of non-commutative
manifolds and self-dual configuration on those manifolds. We construct
a non-commutative generalization of the
Schwarzshild black hole.

\section{Born-Infeld system in curved background}

\label{sec:bi}

First let us study  Born-Infeld theory,  the  famous non-linear
extension of the Maxwell theory. This theory exhibits many
 notable
properties such as duality invariance. This property becomes a guiding
principle for  constructing  other non-linear extensions of non-linear
electrodynamics \cite{rasheed1}. Some general argument for the form of 
non-linear electrodynamics in curved background will be presented in
later sections.
\subsection{Self-duality and topological bound }

In this subsection we shall give a topological bound for Euclidean
Born-Infeld action on a Riemannian manifold and show that it is attained
only by the (anti-)self-dual gauge fields.  
If $({\cal M}, g_{\mu\nu})$ is a Riemannian 4-manifold, then the action
of the Born-Infeld system in this background manifold is 
\begin{eqnarray}
  I(F) = \int_{\cal M} d^4x
  \left(
    \sqrt{\det (g_{\mu\nu} + {\cal F}_{\mu\nu})} - \sqrt{g}
  \right)
\label{actionbi}
\end{eqnarray}
where we have defined $g \equiv \det g_{\mu\nu}$, and 
\begin{eqnarray}
  {\cal F}_{\mu\nu} \equiv F_{\mu\nu} + b_{\mu\nu}.
\end{eqnarray}
Here $b$ is a background two-form which is not necessarily constant,
but we assume that 
\begin{eqnarray}
  d {\cal F} =0.
\end{eqnarray}
Noting that from the (anti-)symmetric property of the indices of the
tensors, we have $\det (g_{\mu\nu}+{\cal F}_{\mu\nu}) = \det
(g_{\mu\nu}-{\cal F}_{\mu\nu})$, thus 
\begin{eqnarray}
\left[\det (g_{\mu\nu}+{\cal F}_{\mu\nu})\right]^2 =
\det (g_{\mu\nu}+{\cal F}_{\mu\rho}{\cal F}_\nu^{\;\;\rho}) g.
\end{eqnarray}
Using Minkowski's inequality\footnote{ 
This inequality can be easily verified in a local orthonormal frame.}
\begin{eqnarray}
  \left[
    \det
    \left(
      g_{\mu\nu}+{\cal F}_{\mu\rho}{\cal F}_\nu^{\;\;\rho}
    \right)
  \right]^{1/4}
\geq 
  \left[
    \det
    \left(
      g_{\mu\nu}
    \right)
  \right]^{1/4}
+
  \left[
    \det
    \left(
      {\cal F}_{\mu\rho}{\cal F}_\nu^{\;\;\rho}
    \right)
  \right]^{1/4},
\end{eqnarray}
we obtain a bound for the action (\ref{actionbi}) as 
\begin{eqnarray}
  I(F) \geq
  \left|
    \int_{\cal M}d^4x \sqrt{\det {\cal F}_{\mu\nu}}
\right|.
\label{boundbi}
\end{eqnarray}
The quantity appearing in the right hand side is a topological
invariant since 
\begin{eqnarray}
  \sqrt{\det {\cal F}_{\mu\nu}} =
\frac14  \left|\sqrt{g}
    {\cal F}_{\mu\nu} * {\cal F}^{\mu\nu}
  \right|,
\end{eqnarray}
where we have defined the Hodge dual of the field strength as
\begin{eqnarray}
  *{\cal F}^{\mu\nu} \equiv \Half \eta^{\mu\nu\rho\sigma}
{\cal F}_{\rho\sigma}. 
\end{eqnarray}
Here $\eta$ is a covariant antisymmetric tensor. 
The equality in (\ref{boundbi}) holds if $g_{\mu\nu} \propto
{\cal F}_{\mu\rho}{\cal F}_\nu^{\;\;\rho}$, and it is easy to show
that this relation results in the (anti-)self-duality condition
\begin{eqnarray}
  {\cal F}_{\mu\nu} = \pm * {\cal F}_{\mu\nu}.
\end{eqnarray}

 From the above argument we observe that in a general curved background
metric the Born-Infeld action is bounded by a topological quantity, 
and the bound is realized when the gauge field configuration is
(anti-)self-dual.

\subsection{The energy momentum tensor}

The above argument supposes a given background metric, however if one
considers a combined system of gravity and a gauge 
field then the gauge field will, in general, affect  the
metric. Here we shall show that in the case of Born-Infeld theory  
self-dual configuration (which we have shown to give a topological
bound for the action) does not affect the background.

By differentiating the action (\ref{actionbi})
with respect to the metric, it is easy to obtain the energy
momentum tensor for this system :  
\begin{eqnarray}
  T^{\mu\nu} =
\frac14 \sqrt{\det (g_{\mu\nu}+{\cal F}_{\mu\nu})}
\left(
(g_{\mu\nu}+{\cal F}_{\mu\nu})^{-1} 
+ (g_{\mu\nu}-{\cal F}_{\mu\nu})^{-1} 
\right)
-
\Half \sqrt{\det g_{\mu\nu}} g^{\mu\nu}.
\end{eqnarray}
Using the technique used in the previous subsection, this expression
can be cast into  the form 
\begin{eqnarray}
  T^{\mu\nu} =
\Half\left[ \det (g_{\mu\nu}+{\cal F}_{\mu\rho}
{\cal F}_\nu^{\;\;\rho})\right]^{1/4}
(\det g_{\mu\nu})^{1/4}
(g_{\mu\nu}+{\cal F}_{\mu\rho}{\cal F}_\nu^{\;\;\rho})^{-1}
-
\Half \sqrt{\det g_{\mu\nu}}g^{\mu\nu}.
\end{eqnarray}
Now, for the self-dual configuration which saturates  Minkowski's
inequality, the gauge field configuration satisfies\footnote{
This equality can be verified easily in local othonormal  frame.}
\begin{eqnarray}
{\cal F}_{\mu\rho}{\cal F}_\nu^{\;\;\rho} = { 1\over 4} 
{\cal F}_{\alpha \beta}{\cal F}^{\alpha \beta}g_{\mu\nu}. 
\label{interes}
\end{eqnarray}
Substituting this condition into the expression for the
energy momentum tensor above, we obtain 
\begin{eqnarray}
  T^{\mu\nu} = 0.
\end{eqnarray}
This shows that in the combined system 
\begin{eqnarray}
  I = I(F) + \int_{\cal M} d^4x \sqrt{g} R(g) 
\end{eqnarray}
this action is extremized on  an Einstein
manifold with an (anti-)self-dual gauge field configuration.
The Einstein equation is not modified by the presence of the matter
gauge field. The question of whether we have a true lower bound is 
slightly delicate. Because the Einstein action is not bounded below,
the solution is only an extremum, not a minimum. However with respect
to  variations
of the gauge field, keeping the metric and the topological term fixed,
it is an absolute minimum.  If we had in mind the idea that
the background metric was a K\"ahler metric, then in fact the
Einstein action when restricted to K\"ahler metrics is purely
topological \cite{Nitku}, once the volume has been fixed. 

The fact that any instanton configuration does not affect the
background metric can be seen also in the context of string theory. In 
equations of motion of supergravity, under a particular ansatz of the
dilaton and R-R scalar field concerning self-dual point, the energy
momentum tensor of them vanishes \cite{gibbonsgreen, chu, bianchi}. 
Thus this D-instantons does 
not affect the Einstein equation. Inclusion of the $b$-field was
discussed in Ref.\ \cite{das}.


\subsection{Open and closed string metric}

Let us comment on the meaning of the background $b$-field. Since we
are considering curved backgrounds, this $b$-field is expected to be
non-constant, under some physical circumstance such as the backgrounds
preserving some of the supersymmetries. So in the following study, the 
$b$-field is not necessarily a constant. 

Following the argument of Ref.\ \cite{SW}, we can define the ``open
string metric'' $G$ as
\begin{eqnarray}
  G_{\mu\nu} \equiv g_{\mu\nu} - B_{\mu\rho} g^{\rho\sigma}
  B_{\sigma\nu}. 
\label{defopen}
\end{eqnarray}
This is a simple generalization of the one given in Ref.\ \cite{SW} in 
which only the constant metric and the constant $b$-field were
considered.
Note that the additional term in eq.\ (\ref{defopen}) contributes
positively so that both $g$ and $G$ are positive definite.
A simple choice of the background $b$-field is one which is
(anti-)self-dual with respect to the ``closed string metric''
$g_{\mu\nu}$. Then from  similar argument to those
presented around eq.\
(\ref{interes}), one can see that $g^{-1}Bg^{-1}B$ is proportional to
the identity matrix. Using the definition (\ref{defopen}), we obtain
an interesting result : the open string metric is conformally
equivalent to the closed string metric,
\begin{eqnarray}
  g_{\mu\nu} \propto G_{\mu\nu}.
\end{eqnarray}

As is well known, the notion of (anti-)self-duality is conformally
invariant :
\begin{eqnarray}
  F_{\mu\nu} = \pm \Half g_{\mu\rho} g_{\nu\sigma}
\eta^{\rho\sigma\tau\lambda} F_{\tau\lambda}
\end{eqnarray}
is invariant under a Weyl rescaling of the metric
\begin{eqnarray}
  g_{\mu\nu} \rightarrow   \Omega^2 g_{\mu\nu} ,
\quad
  \eta^{\mu\nu\rho\sigma} \rightarrow   \Omega^{-4}
  \eta^{\mu\nu\rho\sigma}. 
\end{eqnarray}
So we are lead to the following fact : if $b$ is (anti-)self-dual with
respect to the closed string metric $g$, then the (anti-)self-duality
of $b$ holds also with respect to the open string metric $G$. 

It is easy to see that the inverse of the above claim holds. What one
has to note is that there is a inverse solution of eq.\
(\ref{defopen}) in the matrix notation
\begin{eqnarray}
  g = G + b g^{-1} b = G\left(1 + G^{-1}b\; G^{-1}b + \cdots\right),
\end{eqnarray}
where the ellipsis indicates an infinite power series in
$G^{-1}b\;G^{-1}b$. So if $b$ is (anti-)self-dual with respect to $G$,
the matrix $G^{-1}b\;G^{-1}b$ is in proportion to the identity matrix,
then we see that $g\propto G$ holds. Thus
\begin{eqnarray}
  b = \pm *_g\; b \quad \Leftrightarrow\quad b = \pm 
*_G \; b.
\end{eqnarray}

In conclusion, as long as 
the $b$-field is (anti-)self-dual with respect to closed or
open string metric, both agree about the
(anti-)self-duality of the gauge field configuration. 


\section{General Non-linear Electrodynamics}
\label{sec:abelian}

We now consider a more general type of the non-linear electrodynamics
: 
\begin{eqnarray}
  I(F) = \int \! d^4x \; \sqrt{g} L_F.
\end{eqnarray}
Since we intend to study self-dual configuration in a curved
background, the appropriate property required on the
Lagrangian\footnote{Since we are in the Euclidean regime we use the 
  word Lagrangian to denote the action density. It is the negative of
  the analytic continuation of the usual real time  Lagrangian.} $L_F$ 
is {\it duality invariance}. The duality invariance of general 
non-linear electrodynamics was studied in detail in Ref.\
\cite{rasheed1, Gaillard1, Gaillard2, Gaillard3}, 
in which the authors adopted the following form of the
duality rotation :
\begin{eqnarray}
\left\{
  \begin{array}{l}
  \delta F_{\mu\nu} = * P_{\mu\nu} \\
  \delta P_{\mu\nu} = * F_{\mu\nu} 
  \end{array}
\right.
\label{rotation}
\end{eqnarray}
where the tensor $P$ is defined\footnote{We change $G$ (notation in
  the previous literature) to $P$ because nowadays the former is
  usually taken to denote the open string metric.} 
as 
\begin{eqnarray}
  P^{\mu\nu} \equiv 2 \frac{\partial L_F}{\p F_{\mu\nu}}.
\end{eqnarray}
Note that the duality transformation transforms $F$ into $*P$, not
into $*F$. This is a precise generalization of the electro-magnetic
duality of the Maxwell theory. The duality of this
theory rotates ${\bf E}$ and ${\bf H}$ (or ${\bf D}$ and ${\bf B}$). 
Of course if we adopt the Maxwell Lagrangian in vacuum, $L_F =
F_{\mu\nu} F^{\mu\nu} /4$, then we have $P = F$. 

\subsection{Conditions concerning the self-dual configurations}

Our aim is to clarify the constraints on the form of the Lagrangian
$L_F$ by requiring some physical properties concerning the self-dual
configurations. 
The first requirement is that the well-known (anti-)self-duality
condition  
\begin{eqnarray}
  F= \epsilon * F
\label{usudu}
\end{eqnarray}
with $\epsilon=\pm 1$ must be reduced to the natural
(anti-)self-duality condition expected 
from the above rotation (\ref{rotation}) : 
\begin{eqnarray}
F = \epsilon *P.  
\label{natudu}
\end{eqnarray}
In fact, we see that any nontrivial solution of the condition
(\ref{natudu}) should also be the solution of (\ref{usudu}) and this
restricts the form of the Lagrangian. 
Let us express the explicit form of the Lagrangian as
\begin{eqnarray}
  L_F = L_F(X,Y)
\label{eq:lfpq}
\end{eqnarray}
where we have defined Lorentz-invariant quantities $
  X \equiv F_{\mu\nu} F^{\mu\nu}/4,
  Y \equiv F_{\mu\nu}* F^{\mu\nu}/4$.
Then from this expression we obtain 
\begin{eqnarray}
  P^{\mu\nu} =  \left(
     F^{\mu\nu} \frac{\p L_F}{\p X} 
     +  *F^{\mu\nu} \frac{\p L_F}{\p Y} 
  \right).
\label{defg}
\end{eqnarray}
Substituting the natural (anti-)self-duality (\ref{natudu}) into the
definition above, we have the equation  
\begin{eqnarray}
  \left(
    \epsilon  -  \frac{\p L_F}{\p Y} 
  \right)
*F = 
\left(
   \frac{\p L_F}{\p X} 
\right) F.
\end{eqnarray}
Assuming that this equation has non-trivial ($F\neq 0$) solution, it
must 
also  be a solutions of the usual self-duality equation (\ref{usudu}),
and so we have a condition on the Lagrangian 
\begin{eqnarray}
  \left[
    1  -  \frac{\p L_F}{\p X} -\epsilon \frac{\p L_F}{\p Y}
  \right]_{X = \epsilon Y} =0.
\end{eqnarray}
Here we have used the fact the self-duality condition
implies $X=\epsilon Y$. In the case of Maxwell Lagrangian $L_F
= X$, the constraint above  is trivially satisfied.

A second physical requirement on the Lagrangian is that the
self-dual solution must be a solution of second order
equations of motion. 
The equation of motion for this non-linear system is 
$  \nabla_\nu P^{\mu\nu}= 0.$ Substituting eq.\ (\ref{defg})
into this equation, then we have 
\begin{eqnarray}
\left[  \frac{\p^2 L_F}{\p X^2}
+ 2\epsilon   \frac{\p^2 L_F}{\p X \p Y}
+  \frac{\p^2 L_F}{\p Y^2} 
\right]_{X=\epsilon Y}
=0.
\end{eqnarray}
When obtaining this equation, we have used the  equation $\nabla_\mu
F^{\mu\nu} =0$ which comes from the Bianchi identity and the
(anti-)self-duality condition (\ref{usudu}).

The final requirement is that as seen in the previous section the
background adopted must not be affected by the (anti-)self-dual
configuration considered. Let us calculate the energy momentum
tensor. Using an expression given in Ref.\ \cite{rasheed3}, 
\begin{eqnarray}
  T_{\mu\nu} = \frac{\sqrt{g}}{2}
\left(
  g_{\mu\nu} L_F - P_\mu^{\;\;\lambda} F_{\nu\lambda}
\right).
\end{eqnarray}
Substituting the explicit expression of $P$ (\ref{defg}) and
noting the  relation $F_\mu^{\;\;\lambda} * F_{\nu\lambda} =
g_{\mu\nu}Y$, for the (anti-)self-duality configuration we have 
\begin{eqnarray}
  T_{\mu\nu}= 
\frac{\sqrt{g}}{2} g_{\mu\nu}
\left[
  L_F - X \frac{\p L_F}{\p X} - Y \frac{\p L_F}{\p Y}
\right]_{X=\epsilon Y}.
\end{eqnarray}
Therefore, the vanishing of the energy-momentum tensor 
will hold if 
\begin{eqnarray}
\left[
  L_F - X \frac{\p L_F}{\p X} - Y \frac{\p L_F}{\p Y}
\right]_{X=\epsilon Y} =0.
\end{eqnarray}

To avoid misunderstanding, we should remind the reader that
not every regular solution of the equations of motion need
(anti-)self-dual. We will give an explicit 
counter-example later.
\subsection{Solution of the physical requirements}

Now we summarize these three requirements on the form of the
Lagrangian 
\begin{eqnarray}
  \mbox{(i)} & \mbox{Natural self-duality holds :} &
  \left[
    1 -  \frac{\p L_F}{\p X} -\epsilon \frac{\p L_F}{\p Y}
  \right]_{X = \epsilon Y} =0.\\
  \mbox{(ii)} & \mbox{ Consistency with EOM :} &
  \left[  \frac{\p^2 L_F}{\p X^2}
    + 2\epsilon  \frac{\p^2 L_F}{\p X \p Y}
    +  \frac{\p^2 L_F}{\p Y^2} 
  \right]_{X=\epsilon Y}
  =0. \\
  \mbox{(iii)} & \mbox{Stress tensor vanishes :} &
  \left[
  L_F - X \frac{\p L_F}{\p X} - Y \frac{\p L_F}{\p Y}
  \right]_{X=\epsilon Y} =0.
\end{eqnarray}
At first glance, these three conditions appear to be independent.
However, in the following we show that the three conditions  are
equivalent with each other. Before considering the form of the
solutions of the above conditions, we introduce two useful variables
\begin{eqnarray}
  u \equiv (X + \epsilon Y)/2, \quad v \equiv (X-\epsilon Y)/2.
\end{eqnarray}
Using these variables, the above conditions are written as 
\begin{eqnarray}
  \mbox{(i)} & 
\displaystyle
\left.
\frac{\p L_F}{\p u}
\right|_{v=0} = 1.\\
  \mbox{(ii)} & 
\displaystyle
\left.
\frac{\p^2 L_F}{\p u^2}
\right|_{v=0} = 0.\\
  \mbox{(iii)} & 
\displaystyle
\left[L_F - u
\frac{\p L_F}{\p u}
\right]_{v=0} = 0.
\end{eqnarray}
The Lagrangian $L_F$ may be written as a sum of homogeneous
 polynomials of $u$ and $v$, so
let us adopt the following expression for the Lagrangian :
\begin{eqnarray}
  L_F = \sum_{m,n\geq 0} a_{mn} u^m v^n.
\end{eqnarray}
Now we assume that there is no cosmological term, $a_{00}=0.$
Furthermore, in the weak field approximation, the Lagrangian should be
reduced to the Maxwell Lagrangian, thus
\begin{eqnarray}
  L_F = X + {\cal O} (X^2, XY, Y^2).
\end{eqnarray}
This implies that $a_{01} = a_{10} = 1$. Under these assumptions, we
can express the above three conditions in terms of the coefficient
$a_{mn}$. It is easy to see that all of these conditions are
equivalent with the following constraint :
\begin{eqnarray}
  a_{m0} = 0 \quad {\rm for } \quad m\geq 2.
\label{consa}
\end{eqnarray}
Therefore we conclude that the above three conditions (i) --- (iii)
are equivalent with each other, and give the constraint (\ref{consa})
for the coefficients of the expansion of the Lagrangian.

 From a  physical point of view, perhaps the simplest
 choice for the action is
the one which gives duality-invariant equations of motion
\cite{rasheed1, Gaillard1, Gaillard2, Gaillard3}. Let us consider the
 compatibility of the $SO(1,1)$ 
duality\footnote{The duality group is now
  $SO(1,1)$, not $SO(2)$. This is due to the fact that we are in
  Euclidean regime. See Sec. \ref{sec:hami}.}
invariance requirement and the constraint above (\ref{consa}). 
The condition obtained in Ref.\
\cite{rasheed1} is 
\begin{eqnarray}
  F_{\mu\nu} * F^{\mu\nu} =
  P_{\mu\nu} * P^{\mu\nu}. 
\end{eqnarray}
When this condition is satisfied, the Euler-Lagrange equations for the
gauge field are duality-invariant, and furthermore, the
energy-momentum tensor is also duality-invariant. Thus the equation of
motion for the gravity is also duality-invariant. In terms of our
notation, the above condition is written as 
\begin{eqnarray}
  \mbox{(iv)} & \mbox{Duality invariance :} &
  2 X \frac{\p L_F}{\p X}\frac{\p L_F}{\p Y}
+Y
\left\{
  \left(\frac{\p L_F}{\p X}\right)^2\!  +\! 
\left(\frac{\p L_F}{\p Y}\right)^2 
\right\}
=Y.
\end{eqnarray}
Using $u$ and $v$, this condition is simply expressed as 
\begin{eqnarray}
  \mbox{(iv)} & 
  \displaystyle
  u
  \left(
    \frac{\p L_F}{\p u}
  \right)^2
  -
  v
  \left(
    \frac{\p L_F}{\p v}
  \right)^2
  =u-v.
\label{dualinv}
\end{eqnarray}
Of course the Maxwell Lagrangian $L_F = u+v$ satisfies this
condition. 

Putting $v=0$ in this equation (\ref{dualinv}), then one observes that 
the condition (i) is deduced. Under some plausible assumption we have
seen that three (i) --- (iii) conditions are equivalent, thus
consequently, we have shown that the duality invariant condition (iv)
is sufficient for showing (i) --- (iii). In sum, our conclusion is :
{\it Three conditions ((i) Validity of natural duality, (ii)
  Consistency of self-duality with EOM, (iii) vanishing of stress
  tensor) are equivalent with each other, and any duality invariant
  system possesses all of these three properties.  } 

As discussed in Ref.\ \cite{rasheed1}, it is difficult to solve the
condition (\ref{dualinv}) explicitly. For an example of duality
invariant system, let us consider the Born-Infeld Lagrangian
\begin{eqnarray}
  L_F = \frac{1}{\sqrt{g}}
\sqrt{\det(g_{\mu\nu} + F_{\mu\nu})} -1 
= \sqrt{1 + 2(u+v) + (u-v)^2}-1.
\end{eqnarray}
It is very easy to show that this Lagrangian satisfies the above
condition (\ref{dualinv}), and thus the results of
the previous section are consistent with the argument presented in
this section. See Ref.\ \cite{Hatsuda} for some other solutions.


\section{Higher-rank Generalization}

\label{sec:higher}

In the previous sections, we have seen some interesting properties of
the self-dual configurations and their relation to the
duality-invariant system. Now in this section, we treat the case in
which there are several $U(1)$ gauge fields. As for non-Abelian gauge
groups, comments will be given in the discussion. 

\subsection{Three conditions and their compatibility}

Among many possibilities for the Lagrangian of $n$ gauge fields
$A^{(i)}_\mu$ where $i = 1,\cdots,n$, we adopt the following form of
the Lagrangian 
\begin{eqnarray}
  L_F = L_F(X_i, Y_i)
\end{eqnarray}
which is a simple generalization\footnote{We have assumed that there
  is no cross term such as $F_{\mu\nu}^{(1)}*F^{(2)\mu\nu}$. } of
(\ref{eq:lfpq}). Here the Lorentz-invariant quantities are defined as  
\begin{eqnarray}
X_i \equiv   F_{\mu\nu}^{(i)}F^{(i)\mu\nu}/4 ,\quad
Y_i \equiv   F_{\mu\nu}^{(i)}*F^{(i)\mu\nu}/4. 
\end{eqnarray}
The dual field strength is also generalized as 
\begin{eqnarray}
  P^{(i)\mu\nu} \equiv 2 \frac{\p L_F}{\p F^{(i)\mu\nu}}
=
\left(
  F^{(i)\mu\nu} \frac{\p L_F}{\p X_i}
+*F^{(i)\mu\nu} \frac{\p L_F}{\p Y_i}
\right).
\label{alsogene}
\end{eqnarray}

Using this Lagrangian, it is easy to derive the three conditions in
the previous section. The first requirement, which is the consistency
with the natural self-duality condition with the usual self-duality
condition, is expressed by
\begin{eqnarray}
  \left[
    1 - \frac{\p L_F}{\p X_i}- \epsilon_i \frac{\p L_F}{\p Y_i}
  \right]_{X_j = \epsilon_j Y_j}=0.
\end{eqnarray}
In this expression, we have used a constant $\epsilon_i$ which equals
to $\pm 1$, depending on whether the gauge field $A_\mu^{(i)}$ 
is self-dual or
anti-self-dual. This condition must be understood to hold for
arbitrary $i$. The second requirement of the consistency with
equations of motion reads 
\begin{eqnarray}
\left[
  \frac{\p^2 L_F}{\p X_i \p X_j}
 + \epsilon_j   \frac{\p^2 L_F}{\p X_i \p Y_j}
 + \epsilon_i   \frac{\p^2 L_F}{\p Y_i \p X_j}
 + \epsilon_i \epsilon_j  \frac{\p^2 L_F}{\p Y_i \p Y_j}
\right]_{X_k = \epsilon_k Y_k \;{\rm for}\;{\rm any} k }=0,
\end{eqnarray}
for arbitrary $i$ and $j$. The third condition of the vanishing of the
energy-momentum tensor is 
\begin{eqnarray}
\left[
  L_F - \sum_i
  \left(
  X_i \frac{\p L_F}{\p X_i} +Y_i \frac{\p L_F}{\p Y_i}
  \right)
\right]_{X_j = \epsilon_j Y_j}
=0.
\label{eq:thirdg}
\end{eqnarray}
Let us make a change of variables in a manner similar to the previous
section as 
\begin{eqnarray}
  u_i \equiv (X_i + \epsilon_i Y_i)/2,\quad
  v_i \equiv (X_i - \epsilon_i Y_i)/2.
\end{eqnarray}
Then the three conditions become
\begin{eqnarray}
  \mbox{(i)} & 
\displaystyle
\left.
\frac{\p L_F}{\p u_i}
\right|_{v_j=0} = 1.
\label{eq:threecog1}\\
  \mbox{(ii)} & 
\displaystyle
\left.
\frac{\p^2 L_F}{\p u_i \p u_j}
\right|_{v_k=0} = 0.
\label{eq:threecog2}\\
  \mbox{(iii)} & 
\displaystyle
\left[L_F - \sum_i u_i
\frac{\p L_F}{\p u_i}
\right]_{v_j=0} = 0.
\label{eq:threecog3}
\end{eqnarray}
How these three conditions are compatible? Noting that especially the
third condition (\ref{eq:thirdg}) is only a single condition whereas
the first and the second conditions are more than that, it is obvious
that these three conditions are not equivalent with each other. 
This is in contrast with the case of a single gauge field. However,
as we shall see, condition (i) is equivalent to condition
(ii). Let us assume 
that the Lagrangian is a polynomial of $u_i$ and $v_i$ :
\begin{eqnarray}
  L_F = \sum_{k_1k_2\cdots l_n}a_{k_1k_2\cdots l_n} 
\Pi_{i=1}^{n}u_i^{k_i}\Pi_{i=1}^{n}v_i^{l_i}.
\end{eqnarray}
Then the condition (i) implies 
\begin{eqnarray}
  L_F = \sum_i (u_i+v_i) + O(uv).
\label{eq:formwe}
\end{eqnarray}
Here of course we assumed that the cosmological constant
is zero. This behavior is consistent with the requirement that the
Lagrangian must reduce to the Maxwell system in the weak field limit. 
It is easy to see that eq.\ (\ref{eq:formwe}) is equivalent with
the condition (ii) if we adopt the form of the weak field limit.
The condition (iii) is, however, a weaker one compared to (i) and
(ii). A simple calculation show that the form (\ref{eq:formwe})
satisfies the condition (iii). Thus, we conclude that, under the 
appropriate weak-field assumption, 
\begin{eqnarray}
  {\rm (i) } \Leftrightarrow {\rm (ii)} \Rightarrow {\rm (iii)}.
\end{eqnarray}

\subsection{Duality-invariant systems}

\label{sec:non}

The simple choice for a Lagrangian is the one which gives
duality-invariant equations of motion as seen in the previous
section. Now we have several gauge fields, therefore there are various 
versions of duality rotations. First, let us generalize the argument
given in Ref.\ \cite{rasheed1} and Refs.\ \cite{Gaillard1, Gaillard2}.

The most general duality invariance is defined as \cite{Gaillard3}
\begin{eqnarray}
  \delta
  \left(
    \begin{array}{c}
     *P \\ F
    \end{array}
  \right)
=
  \left(
    \begin{array}{cc}
     A & B \\ C & D
    \end{array}
  \right)
  \left(
    \begin{array}{c}
     *P \\ F
    \end{array}
  \right)
\end{eqnarray}
where $A, B, C, D$ are $n \times n$ matrices. Only when we include
scalar fields which transform appropriately under the duality, the
full duality group $Sp(2n, {\bf R})$ is obtained. However we are
studying a system with only the gauge fields, the full available 
duality group is $U(n)$ which is the maximal compact subgroup of
$Sp(2n, {\bf R})$ (see Sec.\ \ref{sec:adding} for 
the argument with adding scalar fields and enhancement of the
symmetry). This is obtained by imposing  
\begin{eqnarray}
A=D=-A^T,\quad B=-C=B^T.  
\end{eqnarray}
(In this subsection we are temporarily working in the Lorentzian
signature adopted in Ref.\ \cite{rasheed1}.)

Let us consider the general constraint on the constitutive relation 
(\ref{alsogene}) under the requirement of this duality rotation
invariance, following Ref.\ \cite{rasheed1}.
Performing the duality rotation in the both sides of eq.\
(\ref{alsogene}), one has
\begin{eqnarray}
  2A^{ik} \frac{\p L}{\p F^{(k)\mu\nu}}
+
B^{ik}\!*\!  F^{(k)\mu\nu}
=
2
\left[
  -2 B^{jm} 
\left(* \frac{\p L}{\p F^{(m)}_{\rho\sigma}} \right)+
  A^{jm}F^{(m)\rho\sigma} 
\right]
 \frac{\p }{\p F^{(j)\rho\sigma}}
\left( \frac{\p L}{\p F^{(i)\mu\nu}}\right).
\end{eqnarray}
If we demand that this equation is satisfied for any anti-symmetric
matrix $A$ and any symmetric matrix $B$, then the full compact duality 
group  $U(n)$ is obtained. Since $A$ and $B$ are independent of each
other, we have 
\begin{eqnarray}
A^{ik} \frac{\p L}{\p F^{(k)\mu\nu}}
=
  A^{jm}F^{(m)\rho\sigma} 
 \frac{\p }{\p F^{(j)\rho\sigma}}
 \frac{\p L}{\p F^{(i)\mu\nu}},
\label{condA}
\end{eqnarray}
and
\begin{eqnarray}
B^{ik}*  F^{(k)\mu\nu}
=
-4 B^{jm} \left(* \frac{\p L}{\p F^{(m)}_{\rho\sigma}}\right)
 \frac{\p }{\p F^{(j)\rho\sigma}}
 \left(\frac{\p L}{\p F^{(i)\mu\nu}}\right).
\label{condB}
\end{eqnarray}

For the case of $n=1$ as in Ref. \cite{rasheed1}, there is no
antisymmetric $1 \times 1$ matrix, so $A=0$ holds. Then the first
condition does not appear. 
For general $n$, it turns out that the first condition (\ref{condA})
is equivalent with  
the invariance of the Lagrangian under the rotation defined by $A$. 
Using a relation 
\begin{eqnarray}
F^{(m)\rho\sigma} 
 \frac{\p }{\p F^{(j)\rho\sigma}}
 \frac{\p L}{\p F^{(i)\mu\nu}}=
 \frac{\p }{\p F^{(i)\mu\nu}}
\left(
F^{(m)\rho\sigma} 
 \frac{\p L}{\p F^{(j)\rho\sigma}}
\right)
-\delta^{im} 
 \frac{\p L}{\p F^{(j)\mu\nu}},
\end{eqnarray}
we see that the condition (\ref{condA}) reduces to\footnote{We ignore
  the integration constant because it will not appear in general.}
\begin{eqnarray}
A^{jm}F^{(m)\rho\sigma} 
 \frac{\p L}{\p F^{(j)\rho\sigma}}=0.
\end{eqnarray}
This is merely a condition of the invariance of the Lagrangian. 
So if one wants to know whether a given Lagrangian has symmetry
generated by the matrix $A$, then one has only to check the invariance 
of the Lagrangian under the transformation $F^{(i)} \rightarrow
A^{ij}F^{(j)}$.
The largest symmetry associated with $A$ is $SO(n)$.

The second condition for $B$ (\ref{condB}) can be integrated in a
similar manner to Ref.\ \cite{rasheed1}, and the result is
\begin{eqnarray}
F^{(i)}_{\mu\nu}B^{ik}*  F^{(k)\mu\nu}
=
- B^{jm} \left(2 \frac{\p L}{\p F^{(m)}_{\rho\sigma}}\right)
 \left(2*\frac{\p L}{\p F^{(i)\rho\sigma}}\right).
\end{eqnarray}
Here we have neglected the integration constant which violates the
invariance of the energy-momentum tensor. This condition is equivalent 
with  
\begin{eqnarray}
B^{ik} 
\left(
F^{(i)}_{\mu\nu}*  F^{(k)\mu\nu}
+P^{(i)}_{\mu\nu}*  P^{(k)\mu\nu}
\right)=0.
\label{ffpp}
\label{dualgene}
\end{eqnarray}
This is actually the generalization of the condition for $n=1$ 
in Refs.\ \cite{rasheed1, Gaillard1, Gaillard2}. The duality group of
a system can be read from this equation by checking which form of $B$
satisfies the equation (\ref{dualgene}) when a certain Lagrangian is
given.  

For example, if we restrict the form of $B$ as 
\begin{eqnarray}
B = 1_{n}
\label{b1n}
\end{eqnarray}
where $1_n$ is a unit $n \times n$ matrix, then this indicates that the
system satisfying the conditions (\ref{ffpp}) and (\ref{b1n})
possesses only the symmetry involved with simultaneous duality
rotation for all the gauge fields. The corresponding condition in the
Euclidean signature is  
\begin{eqnarray}
\sum_i
\left(
  F^{(i)}_{\mu\nu} * F^{(i)\mu\nu} -
  P^{(i)}_{\mu\nu} * P^{(i)\mu\nu} 
\right)=0.
\label{duain}
\end{eqnarray}
We intend to check how the the simultaneous duality invariance
(\ref{duain}), which is expressed as  
\begin{eqnarray}
  \sum_i
  \left[
    u_i
    \left(
      \frac{\p L_F}{\p u_i}
    \right)^2
-    v_i
    \left(
      \frac{\p L_F}{\p v_i}
    \right)^2
-(u_i - v_i)
  \right]=0, 
\end{eqnarray}
is consistently reproducing the three conditions
(\ref{eq:threecog1})--(\ref{eq:threecog3}). However, it is obvious
that this duality invariance condition is not sufficient to reproduce
them. This is because the number of the conditions is not sufficient.

On the other hand, 
we can require a rather high symmetry such as
\begin{eqnarray}
\{B\}=\{{\rm diag}(0,\cdots,0,1,0,\cdots,0)\; |\; i = 1, \cdots, n \},
\label{ithentry}
\end{eqnarray}
where the diagonal matrix denotes the one in which only the $i$-th
diagonal entry is non-vanishing.  Then of course we can derive a much
restrictive condition
\begin{eqnarray}
  F^{(i)}_{\mu\nu} * F^{(i)\mu\nu} -
  P^{(i)}_{\mu\nu} * P^{(i)\mu\nu} 
  =0
\label{duain2}
\end{eqnarray}
for any $i$. 
In this case the duality group from the $B$ part is $(SO(2))^n$. From 
this condition it is possible to derive all the
three conditions (i) -- (iii). 

 From the above argument, we have seen that the full antisymmetric $B$
is not necessary for deriving the three conditions. Only the form
(\ref{ithentry}) is necessary. 


\subsection{Some examples}

In the following, we present explicit examples of the generalization
of the Born-Infeld action to the higher rank case. We see, however,
that an intriguing example (denoted as example 3 below) satisfies
all the three conditions in spite of the fact that the Lagrangian does 
not satisfy the strong condition (\ref{duain2}).


\subsubsection{Example 1 : Effective action of string theory}

\label{sec:ex1}

The Abelian Born-Infeld Lagrangian admits D-brane interpretation in
string theory. This theory describes low energy effective theory on
a single D-brane. ``Low energy'' means the slowly-varying field 
approximation,
where the derivative of the field strength is neglected.
A straightforward generalization of this D-brane action to the
higher rank case may be obtained in the situation where many parallel 
D-branes are considered. Naively in this case the low energy
description is associated with non-Abelian gauge groups. 
One of the proposed non-Abelian non-linear Lagrangian describing this
system is \cite{Tseytlin}
\begin{eqnarray}
  L_F = {\rm STr} \sqrt{\det (g_{\mu\nu} + F_{\mu\nu})} -\sqrt{g}
\end{eqnarray}
where the gauge fields are non-Abelian matrices, and the the symmetric
trace operation is normalized as  
\begin{eqnarray}
{\rm STr}\;  T^a T^b = \delta^{ab}.
\end{eqnarray}
If we simply restrict the configuration of the gauge fields to only
the diagonal entries, then the resultant Lagrangian is the
summation of the Abelian Born-Infeld Lagrangian. This situation can be 
achieved in string theory when the distance between each brane becomes 
large compared to the string scale.
\begin{eqnarray}
  L_F = \sum_{i=1}^{n}
\left( \sqrt{\det \left(g_{\mu\nu} + F_{\mu\nu}^{(i)}\right)}
-\sqrt{g}\right).
\end{eqnarray}
Obviously, this Lagrangian satisfies the duality invariance condition
with (\ref{ithentry}), since the Lagrangian is merely a summation of
the Abelian Born-Infeld Lagrangian. Thus this
Lagrangian has an $(SO(2))^n$ duality group. In other words each
$U(1)$ field may be subjected to an independent $SO(2)$ duality
rotation defined on each D-brane. Of course the Lagrangian satisfies
the three criteria (\ref{eq:threecog1}) --  (\ref{eq:threecog3}).
As for the $A$ part, there remains $S_n$ symmetry permuting the $n$
$U(1)$s. 

To be summarized, this example corresponds to a duality matrix
restricted to 
\begin{eqnarray}
  A=0, \quad 
\{B\}=\{{\rm diag}(0,\cdots,0,1,0,\cdots,0)\; |\; i = 1, \cdots, n \}.
\end{eqnarray}
The duality group is $(SO(2))^n$ but there is no global
continuous symmetry which rotates the label of the gauge fields.

There is a topological bound for this system. The bound is realized
when the gauge fields are (anti-)self-dual, but one can choose 
self-dual or anti-self-dual independently for each index $(i)$. 
However, in order to preserve the target space supersymmetry in the
D-brane sense, the configuration must be all self-dual or all
anti-self-dual. The other configurations may break supersymmetry.


\subsubsection{Example 2 : A frustrated topological bound}

In Ref.\ \cite{hagiwara}, the following Lagrangian was suggested in
the flat space :
\begin{eqnarray}
  L_F = \sqrt{n
\left(n + 
2\sum_i X_i 
 +\sum_i Y_i^2 
\right)}-n.
\end{eqnarray}
This corresponds to taking the symmetrized trace inside the square
root, and thus seems to be unrelated to string theory, as
pointed out in Ref.\ \cite{Brecher}. There is a topological lower
bound for the action but it is easy to see that it can be attained
only when all the $U(1)$ gauge fields are equal to  each other. 
(The system is `frustrated'.) This is 
reflected in the fact that this Lagrangian satisfies none of the three
criteria (\ref{eq:threecog1}) - (\ref{eq:threecog3}). Specifically,  
\begin{eqnarray}
L_F \geq \sqrt{n \sum_i (1\pm Y_i 
)^2} - n.
\label{eq:inequality}
\end{eqnarray}
The bound is saturated if $F^{(i)}_{\mu\nu} = \pm 
*F^{(i)}_{\mu\nu}$ for any $i$, that
is, when the individual $U(1)$ fields are (anti-)self-dual. 
The right hand side of eq.\ (\ref{eq:inequality}) is certainly bigger
than the topological quantity 
\begin{eqnarray}
L_F\geq \pm \sum_i Y_i. 
\label{eq:inequality2}
\end{eqnarray}
To  derive this fact one uses the following inequality
\begin{eqnarray}
  n \sum_i a_i^2 =
\sum_{i<j}(a_i-a_j)^2 + 
  \left(
    \sum_i a_i
  \right)^2
\geq
  \left(
    \sum_i a_i
  \right)^2.
\end{eqnarray}
Thus although the topological inequality (\ref{eq:inequality2})
holds if the $n$ quantities ${\bf B}^{(i)}\cdot{\bf E}^{(i)} $
on the right-hand-side  are unequal,  nevertheless,
 the topological bound (\ref{eq:inequality2})
can be saturated if and only if  
\begin{eqnarray}
F^{(1)} = \pm *F^{(1)}
=F^{(2)} = \pm *F^{(2)}
=F^{(3)} = \pm *F^{(3)}
=\cdots.
\end{eqnarray}
In this sense, the action is bounded below by the action of the
strictly constrained 
configurations. We cannot attain the bound in a component-wise
fashion.


\subsubsection{Example 3 : A new model with large symmetry}

\label{sec:example3}

In this subsection, we introduce a remarkable model with the $SO(2)$
duality group and $SO(n)$ global symmetry. One of the Lagrangians
which have large duality symmetry was constructed by 
Zumino et al.\ \cite{Brace,Zumino}. Our Lagrangian is different from
theirs. We do not know the connection with string theory. 

Let us consider a duality invariant Euclidean action of the
following form : 
\begin{eqnarray}
  L_F && = \sqrt{1+ 2 \sum_i X_i +
          \left(\sum_i Y_i\right)^2}-1 \nn \\
&& = \sqrt{1+ 2 \sum_i (u_i + v_i) + 
  \left(\sum_i u_i - v_i\right)^2}-1.
\label{ex3ac}
\end{eqnarray}
This Lagrangian satisfies the above simultaneous duality invariance
condition (\ref{duain}). This is a simple generalization of the
Abelian Born-Infeld Lagrangian\footnote{This Lagrangian is not the
  well-known non-Abelian 
  Born-Infeld Lagrangian with only the diagonal entries, as seen in
  comparison with the Lagrangian of Sec. \ref{sec:ex1}. }.  The system
described by this Lagrangian possesses a global $SO(n) \times
SO(1,1)_{\rm elemag} $ symmetry. The $SO(n)$ rotates the label of the
gauge field. The $SO(1,1)$ is the simultaneous 
electro-magnetic duality rotation on all the gauge fields. 
In terms of the matrices $A$ and $B$ in the previous subsection, our
action seems to correspond to
\begin{eqnarray}
  \mbox{arbitrary anti-symmetric }\;\; A, \quad B=1_n
\end{eqnarray}
which indicates that we have the duality group $SO(n) \times SO(2)$.  

Since the action (\ref{ex3ac}) does not satisfy a restrictive
duality invariance condition (\ref{duain2}), it is expected that the
Lagrangian does not possess the three properties (i) --
(iii). However, surprisingly, it is a straightforward calculation to
show that this Lagrangian satisfies all the three conditions (i) --
(iii). So this is one non-trivial example which attain many desirable
properties even in the curved background. 


\section{Aspects of general non-linear electrodynamics}

\label{sec:property}

In this section we illustrate various aspects of non-linear
electrodynamics by taking concrete examples including the example 3 of
Sec.\ \ref{sec:example3}. We shall work in a flat manifold for
simplicity in Sec.\ \ref{sec:hami}, \ref{sec:legendre},
\ref{sec:adding}.

\subsection{Hamiltonian for the $SO(n)\times SO(2)$ symmetric system}

\label{sec:hami}

Let us explore the properties of the new model considered above in
example 3. To see the symmetry of the system, the Hamiltonian
formalism is useful.  The Lorentzian Lagrangian is
\begin{eqnarray}
  L_F = 1-\sqrt{1 + 
\sum_i \left(-|{\bf E}^{(i)}|^2+ |{\bf B}^{(i)}|^2\right)
- \left(\sum_i{\bf E}^{(i)} \cdot {\bf B}^{(i)}\right)^2}.
\end{eqnarray}
Performing a Legendre transformation as in Ref.\ \cite{Gibbons1}, 
we can construct the
Hamiltonian as a function of ${\bf B}^{(i)}$ and ${\bf D}^{(i)}$.
We find that 
\begin{eqnarray}
  H=
\sqrt{
\left(1 + \sum_i |{\bf D}^{(i)}|^2 \right)
\left( 1+ \sum_i|{\bf B}^{(i)}|^2\right)
- \left(\sum_i{\bf D}^{(i)} \cdot {\bf B}^{(i)}\right)^2}
-1.
\end{eqnarray}
This expression is manifestly invariant under $SO(2) \times SO(n)$
where the $SO(2)$ rotates ${\bf B}^{(i)}$ into ${\bf D}^{(i)}$.

In the case $n=2$, the Lagrangian may be written in the form
\begin{eqnarray}
  L_F= \sqrt{1 + 2 \alpha - \beta^2} -1,
\end{eqnarray}
where $\alpha \equiv F_{\mu\nu}\bar{F}^{\mu\nu}/4$ and $\beta \equiv
F_{\mu\nu}* \bar{F}^{\mu\nu}/4$. We have introduced two real gauge
fields and combine them to a single complex gauge field. This
Lagrangian is the one given in Refs.\ \cite{Brace,Zumino}. 

Using the method of Ref.\ \cite{rasheed2}, one may extend the compact
$SO(2)$ duality group to the non-compact group $Sp(2, {\bf R}) (=
SL(2,{\bf R}))$ by
adding an axion and dilaton field which will be done in Sec.\
\ref{sec:adding}.

When dealing with time-independent fields, it is convenient to
introduce potentials. We must first perform a Legendre
transformation to get an action in terms of ${\bf E}$ and ${\bf
  H}$, and then substitute ${\bf E} = \nabla \phi$, ${\bf H} =
\nabla\chi$. We then get the $SO(n)\times SO(2)$ invariant energy
functional 
\begin{eqnarray}
  \hat{L}_F = \sqrt{\left(1 -
\sum_i |\nabla \phi^{(i)}|^2\right)
\left(1- \sum_i|\nabla\chi^{(i)}|^2\right)
- \left(\sum_i\nabla\phi^{(i)} \cdot \nabla\chi^{(i)}\right)^2}
-1.
\end{eqnarray} 
The dimensional reduction is a useful method to realize the duality
symmetry of the system. Surprisingly, this dimensional reduction can
be performed even in the curved space, which will be studied in Sec.\
\ref{sec:reduction}.


\subsection{\bf Legendre self-duality} 

\label{sec:legendre}

The $SO(2)$-invariance of the equations of motion
does not mean  that the Lagrangian is invariant
but it does mean that the quantity
\ben
L+{ 1\over 4}\sum  P^{(i)\mu \nu}F^{(i)}_{\mu \nu}
\een
is invariant, with
\ben
P^{(i)\mu \nu}=-2 {\partial L \over \partial F^{(i)}_{\mu \nu}} 
\een
(we are using conventions appropriate for Minkowski signature).
As pointed out by Gaillard and Zumino
\cite{Gaillard1} 
this gives rise to a number of identities including
\ben
L(*  P^{(i)\mu\nu})= 
L(F^{(i)}_{\mu\nu})+ { 1\over 2} \sum_i
F^{(i)} _{\mu \nu} {\partial L \over \partial F^{(i)}_{\mu \nu}} 
\label{leg}.
\een
They  also pointed out that
 identity (\ref{leg}) may be interpreted in terms of a 
version of Legendre
duality
interchanging the Bianchi identity $d *  F^{(i)}=0$
and equations of motion $d*  P^{(i)}=0$.
Locally the former is equivalent to  $F=dA^{(i)}$,
while locally the latter is equivalent  to $*  P^{(i)}=dB^{(i)}$.
Assuming that the Bianchi identity is given,
the field equation may  obtained by varying
the Lagrangian considered as a function of $dA$, $L=L(dA)$.
Dually one could assume the field equation and obtain
the Bianchi identity from a dual Lagrangian $L_{\rm D}(dB)$.

The two variational principles 
and the two Lagrangians
are related by a Legendre transformation, as
can be seen by passing to first order formalism. 
One considers  the Lagrangian
\ben
L(F, B)=L(F^{(i)})- { 1\over 2} \sum_i *  F^{(i)\mu \nu} 
(\partial _\mu B^{(i)}_\nu - \partial _\nu B^{(i)} _\mu ) 
\label{first}. \een
Variation with respect to $B^{(i)}_\mu$ gives the Bianchi identity
$d*  F=0$. Variation with respect $F^{(i)}_{\mu \nu}$
gives the field equation $*  P^{(i)}=dB^{(i)}$.
Eliminating $F^{(i)}$ in favor of $P^{(i)}$ 
gives $L_{\rm D}(dB^{(i)})$.  Thus
\ben
L_{\rm D}(*  P^{(i)})= L(F^{(i)})+{ 1 \over 2}\sum_i
  P^{(i)\mu \nu} F^{(i)}_{\mu \nu},  
\een
in other words
\ben
L_{\rm D}(*  P^{(i)})=-{\hat L}(P^{(i)\mu \nu}),
\een
where ${\hat L}(P^{(i)})$ is the Legendre transform of $L(F^{(i)})$
\ben
{\hat L}(P^{(i)})=-{ 1\over 2} \sum P^{(i)\mu \nu} F^{(i)}_{\mu \nu}
-L(F^{(i)}) 
\een
such that
\ben
F^{(i)}_{\mu \nu}=-2 {\partial {\hat L}  \over \partial 
P^{(i)\mu \nu}}.\label{inv}
\een
Of course one could have started
with the first order system
\ben
L_{\rm D}(*  P^{(i)} , A)= L_{\rm D}(*  P^{(i)}) 
-{ 1\over 2} \sum *  P^{(i)\mu \nu}
 (\partial _ \mu
{A^{(i)}}_\nu-\partial _\nu {A^{(i)}} _\mu).
\een
Variation with respect to $A^{(i)}_\mu$ yields
the field equation $d*  P^{(i)}=0$, and variation
with respect to $*  P^{(i)}$, using (\ref{inv})
gives the Bianchi identity $d*  F^{(i)}=0$. 

For a general non-linear electrodynamic Lagrangian
the two Lagrangians  $L_{\rm D}(*  P^{(i)}) $ and $L(F^{(i)})$
are unrelated, but for a theory invariant
under $SO(2)$ duality rotations (\ref{leg})
implies that
\ben
L_{\rm D}(*  P^{(i)})= L(*  P^{(i)}).
\een      
 
In our case of example 3, 
we have $*  P^{(i)}=({\bf H}^{(i)}, -{\bf D}^{(i)})$, and the system
has $SO(2)$ invariance as seen before, then the
Legendre-transformed Lagrangian is in the same form as
\ben
{\hat L}= \sqrt{ 1-\sum (|{\bf H}^{(i)}|^2 -|{\bf D}^{(i)}|^2)  
-\Bigl (\sum {\bf H}^{(i)} \cdot {\bf D}^{(i)} \Bigr )^2 }-1. 
\een
Technically this property of the form of the Lagrangian is understood
as follows. First we have Lagrangian $L({\bf E}, {\bf B})$. From this
Lagrangian we obtain Hamiltonian $H({\bf D}, {\bf B})$ by Legendre
transformation. Further dualizing ${\bf B}$ into ${\bf H}$, then one
has perfectly dual description $\hat{L}({\bf D}, {\bf H})$. Now, if
the system has $SO(2)$ duality invariance, then the Hamiltonian is
invariant under the exchange ${\bf D} \leftrightarrow {\bf B}$. So the 
second Legendre transformation is precisely the same calculation of
the first Legendre transformation, and one gets the same form as the
starting point $L$. 

To complete this section we wish to point out that
Legendre self-duality does not imply the existence of a continuous
$SO(2)$ duality invariance. As a counter-example one may take
the theory originally considered by Born, which has 
$L=1- \sqrt{1+2X}$.
A simple calculation reveals that this theory is Legendre self-dual
but also that it does not satisfy the necessary condition for 
$SO(2)$-invariance.
\subsection{Adding dilaton and axion}

\label{sec:adding}

To couple to an axion $a$ and a dilaton $\Phi$ we follow a general
procedure which we shall illustrate by means of a particular
example (model 3 of Sec.\ \ref{sec:example3}). We choose to work in
the Hamiltonian 
formalism and shall exhibit a manifestly $Sp(2,{\bf R})$-invariant
Hamiltonian. The covariant formalism is equally simple and follows
closely the discussion for a single $U(1)$ given in Ref.\
\cite{rasheed2}.  
We begin with inserting appropriate factors of $e^\Phi$ into our
original action, adding a ``$\theta$ term" coupling to the axion and
further adding an $Sp(2,{\bf R})$-invariant kinetic action for the
axion and dilaton 
\begin{eqnarray}
L= L_{\rm NLE} (e^{\Phi} {\bf E}^{(i)}, 
e^\Phi {\bf B}^{(i)})-a\sum ({\bf E}^{(i)}\cdot {\bf B}^{(i)})
+L_{\rm AD} (\Phi,a).
\end{eqnarray}
The action $L_{\rm NLE}$ is now invariant under dilations of 
${\bf E}^{(i)}$ and ${\bf B}^{(i)}$ provided one compensates with an
appropriate  shift of the dilaton. We are naturally led to  introduce 
calligraphic variables which are invariant under dilations : 
$ {\cal E}^{(i)}\equiv  e^\Phi {\bf E}^{(i)}$ and $ {\cal
  B}^{(i)}\equiv e^\Phi {\bf B}^{(i)}$.

We now calculate the electric inductions ${\bf D}^{(i)}$
and find that they  are changed both by a shift and a scaling
\begin{eqnarray}
{\bf D}^{(i)}= e^\Phi {\bf D }^{(i)}
( {\cal E}^{(i)}, {\cal B}^{(i)})- a{\bf B}^{(i)},\label{induction}
\end{eqnarray}
where $ {\bf D }^{(i)}( {\cal E}^{(i)}, {\cal B}^{(i)})$ is the 
constitutive relation in the absence of the axion and dilaton
but regarded as a function of the calligraphic variables.
Clearly $ {\bf D }^{(i)}
( {\cal E}^{(i)}, {\cal B}^{(i)})$ is invariant under a shift of the
dilaton, since it is only a function of the calligraphic variables
which by construction are invariant
and therefore  $e^\phi {\bf D }^{(i)}
( {\cal E}^{(i)}, {\cal B}^{(i)})$ will take on the same factor as
$e^\Phi$. Since ${\bf B}^{(i)}$ takes on the same factor as
$e^{-\Phi}$ 
we need the axion $a$ to take on the same factor as $e^{2\Phi} $
to make all terms on the  the right hand side of (\ref{induction})
scale as $e^\Phi$. This fixes up to an over all multiple
the axion-dilaton Lagrangian $L_{\rm AD}$.
Now  ${\bf D}^{(i)}$ will scale\footnote{Note that we distinguish
  between ${\bf D}^{(i)}$ and ${\bf D }^{(i)} ({\cal E}^{(i)}, {\cal
    B}^{(i)})$.} 
as $e^\Phi$.  
We therefore define  calligraphic inductions by
\begin{eqnarray}
{\cal D}^{(i)}\equiv  e^{-\Phi} \left(
{\bf D}^{(i)}+a{\bf B}^{(i)}\right).
\end{eqnarray} 
By construction  ${\cal D}^{(i)}$ will be invariant under shifts of
the dilaton.
Moreover under the shift of the axion $a\rightarrow a+\alpha$, 
${\cal D}^{(i)}$ will be also be invariant as long as 
${\bf D}^{(i)} \rightarrow {\bf D}^{(i)}-\alpha {\bf B}^{(i)}$.

Explicitly, in our example 3, 
\begin{eqnarray} 
{\bf D}^{(i)}+a{\bf B}^{(i)} = 
\frac{
e^{2\Phi } {\bf E}^{(i)} + e^{4 \Phi} \sum_j ({\bf E}^{(j)}\cdot
{\bf B}^{(j)}) {\bf B}^{(i)}  
}{
\sqrt{ 1-e^{2\Phi} \sum_k (|{\bf E}^{(k)}|^2 -|{\bf B}^{(k)}|^2) 
 -e^{4\Phi}( \sum_k {\bf E}^{(k)} \cdot {\bf B}^{(k)})^2}
}. 
\end{eqnarray}
Thus 
\begin{eqnarray}
{\cal D}^{(i)}  = 
\frac{
{\cal E} ^{(i)} + \sum_j ( {\cal E}^{(j)}\cdot {\cal B}^{(j)})
  {\cal B}^{(i)}  }
{\sqrt{1-\sum_k (|{\cal E}^{(k)}|^2 - |{\cal B}^{(k)}|^2) 
 -( \sum_k  {\cal E}^{(k)}\cdot {\cal  B}^{(k)})^2 } 
}
\end{eqnarray}
which is identical, in terms of the calligraphic variables, to the 
constitutive relation in the absence of the axion and dilaton.

Now one finds that 
\begin{eqnarray} 
H=H_{\rm NLE} + H_{\rm AD},
\end{eqnarray}
where $H_{\rm AD}$ is the Hamiltonian of the axion and dilaton fields
constructed from $L_{\rm AD}$ and which is manifestly 
$Sp(2,{\bf R})$-invariant. And in our example
\begin{eqnarray}
H_{\rm NLE}+1=\frac{ 
1+\sum |{\cal B}^{(i)}|^2
}{
\sqrt{
1-\sum_j (|{\cal E}^{(j)}|^2 - |{\cal B}^{(j)}|^2) 
 -( \sum_j {\cal E}^{(j)}\cdot {\cal  B}^{(j)})^2
}}.
\end{eqnarray}
Thus in terms of the calligraphic variables, $H_{\rm NLE}$
has  the same functional form as it does in terms of the 
usual variables in the absence of the axion and dilaton. We may use
the constitutive relation to express $H_{\rm NLE}$ as a function 
of and ${\cal B}^{(i)}$ and ${\cal D}^{(i)}$. It will be identical
to the expression in the absence of the axion and dilaton. In our case 
\begin{eqnarray}
H_{\rm NLE} + 1= \sqrt{ \left(1+ \sum_i |{\cal B}^{(i)}|^2\right) 
\left(1+ \sum_j |{\cal D}^{(j)}|^2\right) -
\left(\sum_k {\cal B}^{(k)}\cdot {\cal D}^{(k)}\right)^2}.
\end{eqnarray}

It is clear that the resulting Hamiltonian is invariant under shifts
of the dilaton and axion, and this is a general result. What about the 
$SO(2)\subset Sp(2,{\bf R})$ duality rotations? 
In principle it is now possible to express $H_{\rm NLE}$ in terms of
the variables ${\bf B}^{(i)}$ and ${\bf D}^{(i)}$ but this is
complicated and unnecessary. Moreover we don't at this stage know
the appropriate transformation rule in terms of these variables.
Of course we do know how $SO(2)$ acts of the axion and dilaton
so that $L_{\rm AD}$ and hence $H_{\rm AD}$ is invariant. Moreover if
the original theory had $H_{\rm NLE}({\bf B}^{(i)}, {\bf D}^{(i)})$
invariant under $SO(2)$  rotations of ${\bf B}^{(i)}$ into 
${\bf D}^{(i)}$ then $H_{\rm NLE}({\cal B}^{(i)}, {\cal D}^{(i)})$
will also be invariant under $SO(2)$ rotations of ${\cal  B}^{(i)}$
into ${\cal D}^{(i)}$. This is certainly the case of our model 3. It
is clear that the action of $SO(2)$ is the standard one on the
calligraphic variables and from this we may work out the
transformation rule on the original  variables.  

One way of viewing this construction is to note that the variables  
${\bf B}^{(i)}$ into ${\bf D}^{(i)}$ are canonically conjugate
and the passage to the variables ${\cal  B}^{(i)}$ into 
${\cal D}^{(i)}$ is a canonical transformation.


\subsection{Dimensional reduction}

\label{sec:reduction}

It is known that in flat space dimensional reduction
of the Born-Infeld action leads to an effective action in which
duality invariance is manifest.
In this subsection we show that the same is true in curved spacetime.\
We assume a stationary Lorentzian metric which locally at least
is defined on $M\equiv {\bf R} \times \Sigma$
and takes the form
\begin{eqnarray}
ds^2= -e^{2U}(dt+\omega_idx^i)^2 + e^{-2U} \gamma _{ij}dx^idx^j
\end{eqnarray} 
and we consider a stationary $U(1)$ field
\begin{eqnarray}
A=\phi(dt+\omega_idx^i) + \hat{A}_idx^i.
\end{eqnarray} From now on we consider all quantities
as living on $\Sigma$ equipped with the metric $\gamma _{ij}$.
Thus all inner products, covariant derivatives etc.\ 
are taken with respect
to the metric $\gamma_{ij}$.
We have
\begin{eqnarray}
F=d\phi \wedge (dt+\omega) + d{\hat A} + \phi d\omega,
\end{eqnarray}
where $\omega \equiv \omega_i dx^i$.
The electric field is 
${\bf E}=- {\bf \nabla } \phi$
and we define the magnetic field by
\begin{eqnarray}
{\bf B}=*_\gamma (d\hat{A} +\phi d\omega).
\end{eqnarray}
Thus
\begin{eqnarray}
\nabla\cdot {\bf B}= {\bf  \nabla} \phi \cdot {\bf \Omega}, 
\label{cond}
\end{eqnarray}
where ${\bf \Omega}=*_\gamma d\omega$.
The two invariants are given as
\begin{eqnarray}
F_{\mu \nu} F^{\mu \nu}=2( -{\bf \nabla} \phi \cdot {\bf \nabla} \phi
+ e^{4U} {\bf B} \cdot {\bf B} )  
\end{eqnarray}
and
\begin{eqnarray}
F_{\mu \nu} * F^{\mu \nu} =4 e^{2U}{{\bf \nabla} \phi}\cdot {\bf B}.
\end{eqnarray}
Using the fact that  $\sqrt{-g}=e^{-2U} \sqrt {\gamma}$ we find that 
\begin{eqnarray}
\sqrt{-g}F_{\mu \nu} * F^{\mu \nu}=4\sqrt{\gamma}{\bf \nabla} 
\phi \cdot {\bf B}= 
\nabla \left(
4\sqrt{\gamma} 
\left( \phi {\nabla\times A} +{ 1\over 2} \phi^2 {\bf \Omega}\right)
\right),
\end{eqnarray}
which is a total derivative as expected.

The Born Infeld Lagrangian is now
\begin{eqnarray}
\sqrt{\gamma} e^{-2U} 
\Bigl ( 1-\sqrt{1  -{\bf \nabla} \phi 
\cdot {\bf \nabla} \phi + e^{4U} {\bf B}
\cdot {\bf B} -( e^{2U}{{\bf \nabla} \phi}\cdot {\bf B})^2  
 } \Bigr ).
\label{redux}
\end{eqnarray}
We must introduce a Lagrange multiplier $\chi$
to enforce the constraint
(\ref{cond}). Thus we add
\begin{eqnarray}
\chi (\nabla\cdot {\bf B}- {\bf \nabla} \phi \cdot {\bf \Omega}) 
\sqrt{\gamma}. 
\end{eqnarray}
Integration by parts and variation with respect to ${\bf B}$
yields 
\begin{eqnarray}
-e^{-2U} {\bf \nabla} \chi= 
{ 
{\bf B} - {\bf \nabla} \phi ({\bf B} \cdot {\bf \nabla} \phi )
 \over 
\sqrt{1  -{\bf \nabla} \phi \cdot {\bf \nabla} \phi + e^{4U} {\bf B}
\cdot {\bf B} -( e^{2U} {\bf \nabla} \phi  \cdot {\bf B} )^2   }
}.
\end{eqnarray}
One now substitutes back into the Lagrangian to get
\begin{eqnarray}
\sqrt{\gamma}
e^{-2U} \Bigl (1-\sqrt{(1-(\nabla \phi )^2) (1-(\nabla \chi) ^2) -
(\nabla \phi \cdot \nabla \chi)^2}
 \Bigr )-\chi {\bf \nabla} \phi \cdot {\bf \Omega}\sqrt{\gamma}.
\end{eqnarray}
The first term is invariant under the rotation $SO(2)$ which rotates
$\phi$ into $\chi$. 
Using the fact that $\nabla\cdot {\bf \Omega}=0$, 
and an integration by parts,
the last term may be 
re-written in the manifestly $SO(2)$-symmetric form
\begin{eqnarray}
{ 1\over 2} (\phi {\bf \nabla} \chi -\chi {\bf \nabla }\phi) 
\cdot {\bf \Omega}.
\end{eqnarray}
So even in the curved background we see that the dimensional reduction
results in a duality-symmetric Lagrangian. This is also the case for
our new model of Sec.\ \ref{sec:example3}.

Note that if we define ${\bf H}=-\nabla \chi$ the square root is the
same as in flat space except that the inner 
product
$\cdot$ is taken in the metric $\gamma_{ij} $.

To obtain the (anti-)self-dual solutions one needs to consider
$t$, $\omega$ and $\phi$ to be pure imaginary. One sets
\begin{eqnarray}
\phi=\pm i \chi.
\end{eqnarray}

In flat space one may also include in the Dirac-Born-Infeld
action  a scalar field $y$ representing
the transverse displacement of the 3-brane,
In the stationary case one finds that the $SO(2)$ duality
symmetry is enhanced to an $SO(2,1)$ \cite{Gibbons1}.
This $SO(2,1)$ symmetry contains an $SO(1,1)$ subgroup
allowing $\phi$ to be boosted into $y$. The fixed points
, i.e. the solutions with $\phi=\pm y$ are BPS
and  correspond to BIon
solutions  representing strings attached to the 3-brane. 
We have included this and repeated the previous steps.
The resultant dualized Lagrangian is written as
\begin{eqnarray}
  \sqrt{\gamma}e^{-2U}
\left(
1-\sqrt{\det
\left( {\bf v}_R \cdot {\bf v}_S - {\cal G}_{RS}
\right)}
\right)
-\chi\nabla\phi\cdot\Omega\sqrt{\gamma},
\end{eqnarray}
where $R, S=1,2,3$ and the vectors ${\bf v}_R$ are defined as
\begin{eqnarray}
  {\bf v}_1 \equiv \nabla\phi,\quad
  {\bf v}_2 \equiv \nabla\chi,\quad
  {\bf v}_3 \equiv e^U\nabla y.
\end{eqnarray}
the metric in the internal space is defined as ${\cal G}={\rm diag}
(-1,-1,1)$. The first term of this action looks invariant under the
rotation in this internal space and seems to have symmetry $SO(2,1)$. 
However, we find that the presence of $e^{U}$ in the definition of
${\bf v}_3$ prohibits us to write the symmetry in terms of scalar
variables, and breaks the $SO(2,1)$ invariance down to the $SO(2)$
duality subgroup. In addition, the second term is invariant under only
the $SO(2)$ group.

One may also couple to gravity. One adds to (\ref{redux}) 
\begin{eqnarray}
\frac14\sqrt{\gamma} \left(R(\gamma)  -2|\nabla U|^2 +\Half e^{4U}
|{\bf \Omega}|^2\right),
\end{eqnarray}
where we are using a notation $4\pi G=1$.
In order to maintain  the constraint  $\nabla\cdot{\bf \Omega}=0$,
one  also adds
\begin{eqnarray}
\Half\sqrt{\gamma} \psi \nabla\cdot{\bf \Omega}, 
\end{eqnarray}
so that there are now two Lagrange multiplier fields $\psi$ and
$\chi$.  
The quantity  $\psi$ is usually called the NUT-potential.
In the absence of matter, elimination leads
to the $Sp(2,{\bf R})$ invariant action
\begin{eqnarray}
\sqrt{\gamma} 
\left(
\frac14 R(\gamma) - \Half|\nabla U|^2 - \Half e^{-4U}|\nabla \psi|^2
\right).
\end{eqnarray}

In the  case of Riemannian, rather than Lorentzian
metrics one considers $\psi$ to be pure imaginary
and the self-dual metrics with triholomorphic Killing vectors
are obtained by setting $e^{2U}=\pm i\psi$.

If one couples Einstein gravity to Maxwell theory
and proceeds with the elimination of the two Lagrange multipliers, 
we obtain
\begin{eqnarray}
  \sqrt{\gamma}
  \left[
\frac14    R(\gamma)
-\Half|\nabla U|^2
-\Half e^{-4U} |\nabla\psi - \phi \nabla\chi + \chi\nabla\phi|^2
+\Half e^{-2U}
\left(
  |\nabla \phi|^2 + |\nabla\chi|^2
\right)
  \right].
\end{eqnarray}
It is well known that the duality group of Maxwell theory $SO(2)$ is
promoted to an $SU(2,1)$ symmetry which includes $SO(2)$. The group
$SU(2,1)$ is what Hull and Townsend \cite{Hull} call the
U-duality group 
in this case. 

If we repeat this procedure for Born-Infeld
one finds the Lagrangian
\begin{eqnarray}
\lefteqn{  \sqrt{\gamma}
  \left[
    \frac14R(\gamma)
-\Half|\nabla U|^2
-
\Half e^{-4U} |\nabla\psi - \phi \nabla\chi+\chi\nabla\phi|^2
\right.}\nn\\
&&\left.
\hs{20mm}
-
e^{-2U}\!\sqrt{(1\!-\! |\nabla \phi|^2)(1\!-\! |\nabla\chi|^2) 
- ( \nabla\phi\!\cdot\!\nabla\chi)^2}+e^{2U}
  \right].
\end{eqnarray}
We observe that the symmetry $SU(2,1)$ is not present.
It seems therefore that if there is some
extension of Born-Infeld containing gravity which is invariant
under the full $SU(2,1)$ U-duality group, 
it is not simply the obvious coupling of Einstein
gravity coupled to Born-Infeld theory just described.


\section{Examples of solutions in curved space}

\label{sec:curved}

If we do not consider the existence of the $b$-field, then our
procedure of Sec.\ \ref{sec:abelian} and \ref{sec:higher} shows that
many examples of self-dual Maxwell fields on curved manifolds studied
in previous literature can survive even when they are generalized to
duality invariant non-linear electrodynamics. Thus in this section we
concentrate on the situation where the $b$-field is turned on.

\subsection{Born-Infeld instantons on curved manifolds with $b$-field}

One motivation for this paper was to consider non-linear
electrodynamics in a constant or almost constant background $b$-field
in curved Riemannian 4-manifold ${\cal M}$, as for example 
in the work of Ref.\
\cite{Braden} where a single Born-Infeld field was considered. From 
our work above, if the theory satisfies the three
requirements (\ref{eq:threecog1}) -- (\ref{eq:threecog3}), to get a
solution, we may take ${\cal F}$ to be a (anti-)self-dual Maxwell
field. Thus from now on we shall take
Born-Infeld Instanton to mean an $L^2$ Maxwell field $F$
of fixed duality together with a $b$ field 
of the same duality which is closed
and covariantly constant or at least covariantly constant
near infinity. 

Another possible motivation is to see to what extent the S-duality
results for Maxwell theory obtained in Ref.\ \cite{Olive} continue to
hold 
in Born-Infeld theory. 

Maxwell theory on Riemannian
four-manifolds has been much studied.
In particular a fair amount is known
about $L^2$ harmonic forms though many mathematical problems remain.
In what follows we  shall focus 
on the problem of finding suitable $b$-fields, quoting the results
about $L^2$ harmonic forms that we  need.

\subsubsection{ Closed manifolds}  If the manifold ${\cal M}$ 
is closed, that
is compact without boundary, $\partial {\cal M}=0$,
and the solution is square integrable then we
are led to standard Hodge-de-Rham theory.
The number of linearly independent (anti-)self-dual Maxwell 
fields is topological and is given by the Betti-numbers
$b^{\pm}_2({\cal M})$.
Note that even if ${\cal M}$ is symplectic, that is even if
 admits a closed
non-degenerate 2-form $\omega$ such that $d\omega=0$,
and we  regard the manifold  ${\cal M}$ as non-commutative
in the sense that we replace the usual commutative product
on smooth  functions by a deformation of the  Poisson algebra
\cite{Kontsevich}, the {\sl topology} of ${\cal M}$ will not change.
If ${\cal M}$ is taken to be K\"ahler then the K\"ahler form
will be of fixed duality. We shall adopt the  convention that
it is self-dual. Because the manifold is K\"ahler 
it will also be covariantly 
constant.
    
Thus in the K\"ahler case it is natural to take,
as suggested in Ref.\ \cite{Braden}   $b$ 
to be proportional to the K\"ahler form $\omega$. 
The Born-Infeld instantons would then be the harmonic forms 
with the same duality which are
not covariantly constant.  A special case arises when
${\cal M}$ is Hyper-K\"ahler. In that case, regardless of 
whether ${\cal M}$ is closed
or compact  we have a three covariantly constant 2-forms
all of which may be taken to be self-dual.
 The only non-flat simply connected example is $K3$ 
in which case the non-constant two forms are all anti-self-dual
and form a 19 dimensional family. Thus there are no Born-Infeld
Instantons on $K3$.

\subsubsection{Hyper-K\"ahler ALE and ALF spaces}

This is closer to the situation envisaged in Ref.\ \cite{Braden}.
In general a compact or non-compact hyper-K\"ahler
manifold admits no $L^2$  harmonic forms with the same duality 
as the K\"ahler
 forms. Thus again we do not expect any Born-Infeld instantons
in this case.
  
In particular in Ref.\ \cite{Braden} a blow up of $R^4/Z_2$ was 
considered.
If the manifold were Hyper-K\"ahler, it would be
the Eguchi-Hanson metric which is one of the family
of metrics admitting a triholomorphic Killing vector 
$\partial \over \partial \tau$. We have
\begin{eqnarray}
ds^2 = V^{-1} (d \tau +{\bf \omega} \cdot
 d{\bf x })^2 + V (d{\bf x})^2 \label{multi},
\end{eqnarray}
with ${\rm curl}\; {\bf \omega} =\nabla V$. We take
\begin{eqnarray}
V=\epsilon + \sum { 1\over |{\bf x}-{\bf x}_i |}.
\end{eqnarray} 
 If $\epsilon =0$ the metric is ALE. If $\epsilon=1$ it is ALF.
To get the Eguchi-Hanson metric we set $\epsilon=0$ and take two
centres. 
The metric is believed to admit just one $L^2 $ Maxwell field
which is anti-self dual. It is associated with the non-trivial
topology and it is easy to write it down explicitly. Strictly speaking 
there seems to be no completely mathematically rigorous
proof, in the manner of Ref.\ \cite{hitchin}
 that  this 2-form  exhausts the $L^2$
cohomology of Eguchi-Hanson  but this conjecture
is supported by index theorem calculations \cite{Gibpope}.
If $\epsilon =0$ and for $k$ centres there is believed to be just a 
$k-1$ dimensional
space of anti-self-dual $ L^2$ harmonic two-forms associated with the
topology. 
The ALF case is very similar, except in addition to the 
anti-self-dual two-forms
expected from the topology, the exact anti-dual two form
\begin{eqnarray}
d \Bigl( V^{-1} (d\tau +{\bf \omega} \cdot d {\bf x}) \Bigr )
\end{eqnarray}
is also in $L^2$. In the case of one centre, i.e. 
for the Taub-NUT solution it is known \cite{hitchin}
that this is the only $L^2$ harmonic two-form.
 Thus again we get no Born-Infeld
instantons in this case.

\subsection{Almost K\"ahler Manifolds}

Rather than taking $b$ to be covariantly constant as it is for a
K\"ahler manifold, one  could also consider an almost K\"ahler
manifold. This has an almost complex structure $J$ such that $J^2=-1$
with respect to which the metric is hermitian, in other words for any
two vectors $X$ and $Y$ 
\begin{eqnarray}
  g(JX, JY) = g(X,Y).
\end{eqnarray}
The associated two form
\begin{eqnarray}
  \omega(X,Y) = g(X,JY)
\end{eqnarray}
is closed but not covariantly constant. However, just as for a
K\"ahler 
manifold,  the magnitude
squared, ${ 1\over 2}\omega_{\alpha \beta} \omega^{\alpha \beta}$,
of the two form $\omega$ is constant.  The difference from a K\"ahler
manifold
is  
that the almost complex structure $J$ is no longer  integrable.

A theorem of Sekigawa 
\cite{Sekigawa} tells us
that if ${\cal M}$ is a compact Einstein manifold with non-negative
scalar curvature then any almost K\"ahler manifold must in fact be
K\"ahler. Perhaps for that  reason, almost-K\"ahler metrics have not
been much studied 
by physicists. However, if ${\cal M}$ is not compact, then there may
exist complete Ricci-flat almost K\"ahler manifolds which are not
K\"ahler. One 
particularly intriguing example, and in fact the only one
known to us,      
 is 
of the form (\ref{multi}) \cite{Nurowski}. It is however incomplete. 
The metric  and some of its properties are
described in detail in Ref.\ \cite{GibbonsRychenkova}. 
In Ref.\ \cite{almost, almost2} 
the uniqueness of this example is discussed.

The metric is obtained by taking the 
harmonic function $V$  to be proportional
to one of the spatial coordinates $z$ say.
Thus it is given by
\begin{eqnarray}
ds^2= z^{-1} (d\tau + xdy)^2 + z( dx^2 +dy^2 +dz^2) \label{Heis}.
\end{eqnarray}
Clearly there is a singularity at $z=0$ but the metric is complete
as $z\rightarrow \infty$. As explained in Ref.\
\cite{GibbonsRychenkova} 
the singularity may be removed by regarding the metric (\ref{Heis})
as the asymptotic form of a complete non-singular hyper-K\"ahler
metric on  
the complement of a smooth cubic curve in $CP^2$. 

The metric (\ref{Heis}) has a Kaluza-Klein interpretation
(reducing on the coordinate $\tau$) as a magnetic
field in the z-direction which is uniform in $x$ and $y$.
Appropriately enough the isometry group
is of Bianchi type II, or  Heisenberg group. The
Maurer-Cartan forms are $(dx,dy, d\tau+xdy)$.
It is well known that the Heisenberg group is the symmetry group
of a particle moving in a uniform magnetic field and
in some circumstances the spatial coordinates
$x$ and $y$ may be regarded as non-commuting. 
Thus it is a very tempting to  speculate
that this metric and its smooth resolution may have something to do
with curved non-commutative spaces. In any event it encourages
its further investigation.

One easily checks that the three (self-dual) K\"ahler forms are
\begin{eqnarray}
dx \wedge(d\tau+xdy) + zdz\wedge dy,\\
dy \wedge(d\tau+xdy) + zdy\wedge dz,
\end{eqnarray}
and 
\begin{eqnarray}
dz \wedge(d\tau +xdy) + zd x\wedge dy.
\end{eqnarray}

In addition there is a circle of anti-self-dual
almost K\"ahler forms
\begin{eqnarray}
\cos \theta \Bigl ( dx \wedge (d\tau+xdy) - zdx \wedge dy \Bigr ) + 
\sin \theta \Bigl ( dy \wedge (d\tau+xdy) - zdy\wedge dy \Bigr ),
 \label{almost}
\end{eqnarray}
where $\theta$ is a constant. The triholomorphic Killing field
$\partial \over \partial \tau$ gives rise to an anti-self dual
Maxwell field of the form
\begin{eqnarray}
 d(z^{-1}(d \tau +x dy) ) \label{max}
\end{eqnarray}
whose magnitude squared is proportional to $z^{-4}$. Since the volume
element 
is $zd\tau dx dy dz$, the $L^2$ norm converges as $z\rightarrow\infty$
but diverges as one approaches the singularity at $z=0$.

An interesting question is whether the almost K\"ahler structure 
(\ref{almost}) or the Maxwell field (\ref{max}) extend
smoothly over the resolution. If the almost K\"ahler structure
extends, it would give a complete  Ricci-flat example of an almost 
K\"ahler  structure which is not K\"ahler.



\subsection{A black hole in an external $b$-field}

In the previous subsection, we showed that ALE and ALF spaces in a
background $b$-field do not admit interesting Born-Infeld instantons. 
The situation changes if we consider Euclidean Schwarzshild solutions.
The metric is
\begin{eqnarray}
  ds^2 = \left(1-\frac{2M}{r}\right)d\tau^2 +
  \left(1-\frac{2M}{r}\right)^{-1} dr^2 + r^2 
  d\Omega^2 .
\end{eqnarray}
 The $L^2$ harmonic two-forms are spanned by \cite{Etesi}
 \begin{eqnarray}
   F_m = \sin \theta d \theta \wedge d \varphi
 \end{eqnarray}
and 
\begin{eqnarray}
  F_e = *F_m=(d \tau   \wedge dr)/r^2.
\end{eqnarray}
Thus $F_m \pm F_e \equiv F_\pm$ is a (anti-)self-dual Maxwell
instanton. To get a 
$b$-field, consider a harmonic two-form $B$ 
\begin{eqnarray}
  B = d
  \left(
    \frac{r^2}{2} \sin^2 \theta d \varphi
  \right).
\end{eqnarray}
The Maxwell field $B$ is neither (anti-)self-dual nor
square-integrable. It looks like a uniform magnetic field near
infinity.  However, 
\begin{eqnarray}
  b_\pm= B\pm *B 
\end{eqnarray}
is (anti-)self-dual and is approximately covariantly constant near
infinity. Thus in general (anti-)self-dual solution of Maxwell
equations in a (anti-)self-dual background $b$-field may be obtained
by taking an arbitrary linear combination of $F_\pm$ and $b_\pm$.
It is amusing to speculate that this solution may have something to do 
with ``non-commutative black holes''.

A smooth square integrable solution of the general non-linear
electrodynamic equations of motion which is not self-dual may be
obtained by taking $P= F_e$. This satisfies the equation of motion
$d * P=0$. Now one may calculate $F$ using the constitutive
relations. This will give something of the form $F=f(r) d\tau \wedge
dr$, for a function $f(r)$ which will depend upon the particular
theory. One has $dF=0$ and therefore it also satisfies the Bianchi
identities.

\subsection{Kaluza-Klein monopoles in constant $b$-fields}

A similar  construction to that of the previous section may  be
applied to axisymmetric Hyper-K\"ahler metrics.
Let us suppose that the metric is of the form (\ref{multi})
but where $V$ and $\bf \omega$ are independent of the angular
coordinate 
$\varphi=\arctan ({y \over x})$. We may choose a gauge in which
\begin{eqnarray}
 ds^2 = V^{-1} (d \tau +\nu d\varphi)^2 + V ( dz^2 + d \rho ^2 
+ \rho d\varphi
^2 ) ,
\end{eqnarray}
where $\rho^2 =x^2 +y^2$ and $V$ and $\nu$ depend only on $\rho$ and
$z$. 

The Killing vector $m^\alpha \partial _\alpha={\partial \over 
\partial \varphi}$ is not triholomorphic, and  the two-form
\begin{eqnarray}
d( g_{\alpha \beta} m^\alpha d x^\beta )  \label{maggie}
\end{eqnarray}
is an anti-self-dual solution of Maxwell's equations.
If the metric is $ALF$ then near infinity  (\ref{maggie}) will look
like a magnetic field and be almost constant. Thus we could take it as
our $b$ field. To get Born-Infeld instantons we can now take
anti-self-dual $L^2$  harmonic two-forms. The simplest case would be
the Taub-NUT metric. It admits a unique \cite{hitchin}  $L^2$
Maxwell field which is anti-self-dual and it will persist as one turns
on the $b-$ field.
  

\section{Conclusion and Discussion}

In this paper, we have analyzed non-linear electrodynamics in curved
space especially from the viewpoint of self-duality and self-dual
configurations. We have observed that various properties of the
self-dual configurations can be derived from the duality invariance of
the system. If the equations of motion are duality invariant, then the
self-dual configurations satisfy the following properties : (i) they
also satisfy the natural duality $F = \pm *P$, (ii) they are also the
solutions of equations of motion, (iii) their stress tensors vanish.

In the $U(1)$ case these three properties are equivalent with a single 
constraint on the form of the Lagrangian. However, if we consider
several gauge fields, then the situation is rather different. We
analyzed the form of the constraint and gave a general argument. 
We presented a new action which possesses $SO(n)\times SO(2)$ duality
symmetry and the above three properties. The extension to include
axion and dilaton gives an enhanced duality group, and this was
checked for our new action. The relation to Legendre duality was
clarified. 

Motivated by the work \cite{Braden}, we studied the inclusion of
self-dual $b$-field background. We gave some explicit examples of
these ``non-commutative manifolds'' and studied self-dual
configurations of the gauge fields on them. 

In the case that the curved background admits a Killing vector
field, we obtained  the dimensional reduction
of the Born-Infeld action and it's coupling to Einstein
gravity. Interestingly, while the $SO(2)$ duality symmetry 
becomes manifest, other symmetries such as the $SO(2,1)$ symmetry
which appears in flat space when one introduces a transverse scalar
or the $SU(2,1)$ U-duality group which emerges when one couples
Einstein gravity to Maxwell theory are broken.
This may indicate that there are other more symmetrical ways of
coupling Born-Infeld theory and gravity consistent with U-duality. 

In this paper we have seen various dualities of non-linear
electrodynamics. We can summarize them in the following table: 

\begin{tabular}{lcc}
 1) Legendre (discrete) self-duality & : & $F \leftrightarrow *P$ \\
 2) $SO(2)$ duality invariance   &   : &Rotation among ($F$, $*P$) \\
 3) $Sp(2,{\bf Z})$ invariance       &   : &Discrete rotation among 
                                          ($F$, $*P$, $\Phi$, $a$)\\
 4) $Sp(2,{\bf R})$ invariance       &   : & Continuous version of 3)
\end{tabular}

\noindent
We have a flowchart representing the relations between the above
dualities as 
\begin{eqnarray}
4) \Rightarrow 3) \Rightarrow 1),\quad
4) \Rightarrow 2) \Rightarrow 1).
  \nonumber
\end{eqnarray}

Some comments concerning the results of this paper are in order.

In sec.\ \ref{sec:higher}, we assumed that Lagrangian is written only
in terms of $X_i$ and $Y_i$. However, in general the combination
$F_{\mu\nu}^{(i)} F^{(j)\mu\nu}$ with $i \neq j$ can appear. Inclusion
of this term gives rise to a problem : we cannot write the
energy-momentum 
tensor in terms of these Lorentz invariant quantities. Since the
general duality condition on the  matrix $B$ (\ref{ffpp}) involves
this cross term in an essential way, a  generalization of our
argument must be provided in order to see the full content of the
structure of the duality groups. 

Our original motivation was to see what sort of Lagrangian is
consistent with  self-dual configurations in a curved
background. This may some insight  into determining the form of
the D-brane action. Note that even in a flat background the precise
form of the non-Abelian D-brane action is not yet known
with certainty. Integrating
out the off-diagonal gauge field components will produce  a
complicated form, but it must be constrained by our duality condition
in the case of many $U(1)$ gauge fields. 

In Ref.\ \cite{Spense},  an Abelian Born-Infeld generalization
of the topological Donaldson-Witten
model was constructed, and it was shown that BRST-invariant
quantities coincide with those of the Maxwell-Donaldson-Witten
topological model. As stated in Ref.\ \cite{Spense},
the explicit form of the Lagrangian was not used,
merely the properties of it's extrema. It seems likely therefore
that topological theories constructed from  other non-linear
electrodynamical theories satisfying our conditions will also have the
same BRST-invariant quantities as Maxwell-Donaldson-Witten theory.
   
We would like to comment on the relation to the recent work on
so-called non-linear BPS equations \cite{SW, nlbps, marino, rot, 
moriyama, moriyama2}. The
configurations considered in this paper are solutions of linear BPS
equations, that is  self-dual equations. But in general under the
presence of a $b$-field, the duality condition may be relaxed because
of the freedom introduced by  the boundary condition of the gauge
field at  spatial infinity. In most previous work only  flat
backgrounds were adopted. In this paper we have shown that the linear
BPS equation survives when we go to a curved background, so it is
expected that the non-linear BPS equation can be derived in a curved 
background. One supporting reason is that the solutions of the
non-linear BPS equations are simply derived from the linear BPS
solution by making use of a target space rotation \cite{rot,
  moriyama, hhm}.  We leave these issues to the future work.



\vs{10mm}
 
\noindent
{\Large\bf Acknowledgment}

G.\ W.\ G.\ would like to thank Professor T.\ Nakamura and other
members of the Yukawa
Institute for their hospitality during his stay in Kyoto
where this work was initiated. He would also like
to thank the organizers of the KIAS Summer
Workshop on Branes
where it continued and Paul Townsend and Malcom Perry
for helpful discussions.
K.\ H.\ would like to thank S.\ R.\ Das and K.\ Itoh for useful
comments. The work of K.\ H.\ is supported in part by Grant-in-Aid for 
Scientific Research from Ministry of Education, Science, Sports and
Culture of Japan (\#02482), and by the Japan Society for the Promotion
of Science under the Postdoctoral Research Program.

\newcommand{\J}[4]{{\sl #1} {\bf #2} (#3) #4}
\newcommand{\andJ}[3]{{\bf #1} (#2) #3}
\newcommand{\AP}{Ann.\ Phys.\ (N.Y.)}
\newcommand{\MPL}{Mod.\ Phys.\ Lett.}
\newcommand{\NP}{Nucl.\ Phys.}
\newcommand{\PL}{Phys.\ Lett.}
\newcommand{\PR}{ Phys.\ Rev.}
\newcommand{\PRL}{Phys.\ Rev.\ Lett.}
\newcommand{\PTP}{Prog.\ Theor.\ Phys.}
\newcommand{\hep}[1]{{\tt hep-th/{#1}}}

\end{document}